\documentstyle[12pt,epsfig]{article}

\catcode`\@=11
\def\@citex[#1]#2{%
\if@filesw \immediate \write \@auxout {\string \citation {#2}}\fi
\@tempcntb\m@ne \let\@h@ld\relax \def\@citea{}%
\@cite{%
  \@for \@citeb:=#2\do {%
    \@ifundefined {b@\@citeb}%
      {\@h@ld\@citea\@tempcntb\m@ne{\bf ?}%
      \@warning {Citation `\@citeb ' on page \thepage \space undefined}}%
      {\@tempcnta\@tempcntb \advance\@tempcnta\@ne%
      \@tempcntb\number\csname b@\@citeb \endcsname \relax%
      \ifnum\@tempcnta=\@tempcntb 
	\ifx\@h@ld\relax%
	  \edef \@h@ld{\@citea\csname b@\@citeb\endcsname}%
	\else%
	  \edef\@h@ld{\ifmmode{-}\else--\fi\csname b@\@citeb\endcsname}%
	\fi%
      \else
	\@h@ld\@citea\csname b@\@citeb \endcsname%
	\let\@h@ld\relax%
      \fi}%
    \def\@citea{,\penalty\@highpenalty\,}%
  }\@h@ld
}{#1}}

\def\@citeb#1#2{{[#1]\if@tempswa , #2\fi}}
\def\@citeu#1#2{{$^{#1}$\if@tempswa , #2\fi }}
\def\@citep#1#2{{#1\if@tempswa , #2\fi}}

\def\bcites{         
	\catcode`\@=11
	\let\@cite=\@citeb
	\catcode`\@=12
}

\def\upcites{         
	\catcode`\@=11
	\let\@cite=\@citeu
	\catcode`\@=12
}

\def\plaincites{      
	\catcode`\@=11
	\let\@cite=\@citep
	\catcode`\@=12
}

\newcount\hour
\newcount\minute
\newtoks\amorpm
\hour=\time\divide\hour by 60
\minute=\time{\multiply\hour by 60 \global\advance\minute by-\hour}
\edef\standardtime{{\ifnum\hour<12 \global\amorpm={am}%
	\else\global\amorpm={pm}\advance\hour by-12 \fi
	\ifnum\hour=0 \hour=12 \fi
	\number\hour:\ifnum\minute<10 0\fi\number\minute\the\amorpm}}
\edef\militarytime{\number\hour:\ifnum\minute<10 0\fi\number\minute}

\def\draftlabel#1{{\@bsphack\if@filesw {\let\thepage\relax
   \xdef\@gtempa{\write\@auxout{\string
      \newlabel{#1}{{\@currentlabel}{\thepage}}}}}\@gtempa
   \if@nobreak \ifvmode\nobreak\fi\fi\fi\@esphack}
	\gdef\@eqnlabel{#1}}
\def\@eqnlabel{}
\def\@vacuum{}
\def\marginnote#1{}
\def\draftmarginnote#1{\marginpar{\raggedright\scriptsize\tt#1}}
\def\draft{
	\pagestyle{plain}
	\overfullrule=2pt
	\oddsidemargin -.5truein
	\def\@oddhead{\sl \phantom{\today\quad\militarytime} \hfil
	\smash{\Large\sl DRAFT} \hfil \today\quad\militarytime}
	\let\@evenhead\@oddhead
	\let\label=\draftlabel
	\let\marginnote=\draftmarginnote
	\def\ps@empty{\let\@mkboth\@gobbletwo
	\def\@oddfoot{\hfil \smash{\Large\sl DRAFT} \hfil}
	\let\@evenfoot\@oddhead}
	\def\@eqnnum{(\theequation)\rlap{\kern\marginparsep\tt\@eqnlabel}%
	\global\let\@eqnlabel\@vacuum}  }

\def\blackfonts{
	\font\blackboard=msbm10 scaled\magstep1
	\font\blackboards=msbm8
	\font\blackboardss=msbm6
}

\def\nblack{            
	\def\ZZ{{Z \n{10} Z}}
	\def\NN{{N \n{14} N}}
	\def\CC{{C \n{11} C}}
	\def\RR{{R \n{11} R}}
	\def\QQ{{Q \n{12} Q}}
	\def\PP{{P \n{11} P}}
}

\def\prep{         
	\catcode`\@=11
	\input art10.sty
	\catcode`\@=12
	
	\let\small\null
	\def\blackfonts{
		\font\blackboard=msbm10
		\font\blackboards=msbm7
		\font\blackboardss=msbm5
	}
	\let\sl\it
	\twocolumn
	\sloppy
	\voffset=-2.54truecm
	\hoffset=-2.54truecm
	\flushbottom
	\parindent 1em
	\leftmargini 2em
	\leftmarginv .5em
	\leftmarginvi .5em
	\marginparwidth 48pt
	\marginparsep 10pt
	\setlength{\columnsep}{2truecm}
	\setlength{\textwidth}{25.4truecm}
	\setlength{\textheight}{17truecm}
	\baselineskip=16pt
	\oddsidemargin .18truein
	\evensidemargin .17truein
}

\def\eqalign#1{\null\,\vcenter{\openup\jot\m@th
  \ialign{\strut\hfil$\displaystyle{##}$&$\displaystyle{{}##}$\hfil
      \crcr#1\crcr}}\,}
\def\eqalignno#1{\displ@y \tabskip\centering
  \halign to\displaywidth{\hfil$\@lign\displaystyle{##}$\tabskip\z@skip
    &$\@lign\displaystyle{{}##}$\hfil\tabskip\centering
    &\llap{$\@lign##$}\tabskip\z@skip\crcr
    #1\crcr}}

\def\section{\@startsection {section}{1}{\z@}{3.ex plus 1ex minus
 .2ex}{2.ex plus .2ex}{\large\bf}}
\def\subsection{\@startsection{subsection}{2}{\z@}{2.75ex plus 1ex minus
 .2ex}{1.5ex plus .2ex}{\bf}}

\def\appendix{{\newpage\section*{Appendix}}\let\appendix\section%
	{\setcounter{section}{0}
	\gdef\thesection{\Alph{section}}}\section}

\def\abstract{\if@twocolumn
\section*{Abstract}
\else 
\begin{center}
{\bf Abstract\vspace{-.5em}\vspace{0pt}}
\end{center}
\quotation
\fi}

\catcode`\@=12

\newcommand{\beq}{\begin{equation}}
\newcommand{\eeq}{\end{equation}}
\newcommand{\beqa}{\begin{eqnarray}}
\newcommand{\eeqa}{\end{eqnarray}}

\newcommand{\Z}{{\bf Z}}

\newcommand{\C}{{\bf C}}
\newcommand{\e}{{\rm e}}

\newcommand{\tilQ}{\widetilde{Q}}
\newcommand{\tilA}{\widetilde{A}}

\def\noj#1,#2,{{\bf #1} (19#2)\ }
\def\jou#1,#2,#3,{{\sl #1\/ }{\bf #2} (19#3)\ }
\def\ann#1,#2,{{\sl Ann.\ Physics\/ }{\bf #1} (19#2)\ }
\def\cmp#1,#2,{{\sl Comm.\ Math.\ Phys.\/ }{\bf #1} (19#2)\ }
\def\ma#1,#2,{{\sl Math.\ Ann.\/ }{\bf #1} (19#2)\ }
\def\jd#1,#2,{{\sl J.\ Diff.\ Geom.\/ }{\bf #1} (19#2)\ }
\def\invm#1,#2,{{\sl Invent.\ Math.\/ }{\bf #1} (19#2)\ }
\def\cq#1,#2,{{\sl Class.\ Quantum Grav.\/ }{\bf #1} (19#2)\ }
\def\cqg#1,#2,{{\sl Class.\ Quantum Grav.\/ }{\bf #1} (19#2)\ }
\def\ijmp#1,#2,{{\sl Int.\ J.\ Mod.\ Phys.\/ }{\bf A#1} (19#2)\ }
\def\jmphy#1,#2,{{\sl J.\ Geom.\ Phys.\/ }{\bf #1} (19#2)\ }
\def\jams#1,#2,{{\sl J.\ Amer.\ Math.\ Soc.\/ }{\bf #1} (19#2)\ }
\def\grg#1,#2,{{\sl Gen.\ Rel.\ Grav.\/ }{\bf #1} (19#2)\ }
\def\mpl#1,#2,{{\sl Mod.\ Phys.\ Lett.\/ }{\bf A#1} (19#2)\ }
\def\nc#1,#2,{{\sl Nuovo Cim.\/ }{\bf #1} (19#2)\ }
\def\np#1,#2,{{\sl Nucl.\ Phys.\/ }{\bf B#1} (19#2)\ }
\def\pl#1,#2,{{\sl Phys.\ Lett.\/ }{\bf #1B} (19#2)\ }
\def\pla#1,#2,{{\sl Phys.\ Lett.\/ }{\bf #1A} (19#2)\ }
\def\pr#1,#2,{{\sl Phys.\ Rev.\/ }{\bf #1} (19#2)\ }
\def\prd#1,#2,{{\sl Phys.\ Rev.\/ }{\bf D#1} (19#2)\ }
\def\prl#1,#2,{{\sl Phys.\ Rev.\ Lett.\/ }{\bf #1} (19#2)\ }
\def\prp#1,#2,{{\sl Phys.\ Rept.\/ }{\bf #1C} (19#2)\ }
\def\ptp#1,#2,{{\sl Prog.\ Theor.\ Phys.\/ }{\bf #1} (19#2)\ }
\def\ptpsup#1,#2,{{\sl Prog.\ Theor.\ Phys.\/ Suppl.\/ }{\bf #1} (19#2)\ }
\def\rmp#1,#2,{{\sl Rev.\ Mod.\ Phys.\/ }{\bf #1} (19#2)\ }
\def\yadfiz#1,#2,#3[#4,#5]{{\sl Yad.\ Fiz.\/ }{\bf #1} (19#2) #3%
\ [{\sl Sov.\ J.\ Nucl.\ Phys.\/ }{\bf #4} (19#2) #5]}
\def\zh#1,#2,#3[#4,#5]{{\sl Zh.\ Exp.\ Theor.\ Fiz.\/ }{\bf #1} (19#2) #3%
\ [{\sl Sov.\ Phys.\ JETP\/ }{\bf #4} (19#2) #5]}

\def\beq{\begin{equation}}
\def\eeq{\end{equation}}
\def\beqar{\begin{eqnarray}}
\def\eeqar{\end{eqnarray}}

\newcommand{\tilq}{\widetilde{q}}
\newcommand{\NS}{{\rm NS}}
\newcommand{\D}{{\rm D}}

\newcommand{\be}{\begin{equation}}
\newcommand{\ee}{\end{equation}}
\newcommand{\bea}{\begin{eqnarray}}
\newcommand{\eea}{\end{eqnarray}}

\def\nfrac#1#2{{\displaystyle{\vphantom1\smash{\lower.5ex\hbox{\small$#1$}}%
	\over\vphantom1\smash{\raise.25ex\hbox{\small$#2$}}}}}

\def\p#1{\mskip#1mu}
\def\n#1{\mskip-#1mu}
\def\stop{\p6.}
\def\comma{\p6,}

\def\to{\rightarrow}

\def\lae{\mathrel{\mathop{\smash{\lower .5 ex \hbox{$\stackrel<\sim$}}}}}
\def\lae{\mathrel{\mathop{\smash{\lower .5 ex \hbox{$\stackrel>\sim$}}}}}

\def\pa{\partial}

\def\l:{\mathopen{:}\,}
\def\r:{\,\mathclose{:}}

\catcode`\@=11
\def\theequation{\arabic{equation}}
\catcode`\@=12

\nblack
\bcites

\nblack

\catcode`\@=11
\def\theequation{\thesection.\arabic{equation}}
\@addtoreset{equation}{section}
\@addtoreset{footnote}{section}
\@addtoreset{footnote}{subsection}
\catcode`\@=12

\newcommand{\beqn}{\begin{equation}}
\newcommand{\eeqn}{\end{equation}}
\newcommand{\beqnarray}{\begin{eqnarray}}
\newcommand{\eeqnarray}{\end{eqnarray}}
\newcommand {\bear} [1] {\begin {array} {#1}}
\newcommand {\ear} {\end {array}}

\newcommand {\beqarn} {\begin{eqnarray*}}
\newcommand {\eeqarn} {\end{eqnarray*}}

\begin{document}
\begin{titlepage}

\begin{center}
\today
\hfill LBNL-039423, UCB-PTH-97/09 \\
\hfill                  hep-th/9702154

\vskip 1.5 cm
{\large \bf Branes and Mirror Symmetry in $N=2$ Supersymmetric 
Gauge Theories in Three Dimensions}
\vskip 1 cm 
{Jan de Boer, Kentaro Hori, Yaron Oz and Zheng Yin}\\
\vskip 0.5cm
{\sl Department of Physics,
University of California at Berkeley\\
366 Le\thinspace Conte Hall, Berkeley, CA 94720-7300, U.S.A.\\
and\\
Theoretical Physics Group, Mail Stop 50A--5101\\
Ernest Orlando Lawrence Berkeley National Laboratory, 
Berkeley, CA 94720, U.S.A.\\}

\end{center}

\vskip 0.5 cm
\begin{abstract}
We use  brane configurations 
and  $SL(2,{\bf Z})$ symmetry of the type IIB string to construct
mirror $N=2$ supersymmetric gauge theories in three dimensions.
The mirror map exchanges Higgs and Coulomb branches, Fayet-Iliopoulos and mass 
parameters and $U(1)_R$
symmetries.
Some quantities that are determined at the quantum level in one theory
are determined at the classical level of the mirror.
One such example is the complex structure
of the Coulomb branch of one theory, which is determined quantum mechanically.  
It is mapped to the complex structure of the Higgs branch of the mirror theory, which is 
determined classically.
We study the generation of $N=2$ superpotentials by open D-string instantons in 
the 
brane configurations.

\end{abstract}

\end{titlepage}

\section{Introduction}
Recently brane configurations that preserve $\frac{1}{8}$ of space-time 
supersymmetry of type II string theory
have been used to study $N=1$ duality in four dimensions \cite{EGK}.
Similar configurations can be used, upon T-dualizing one of the coordinates,
to study $N=2$ supersymmetric gauge theories in three dimensions.
In \cite{si,dhoo,pz,HW,dhooy} a mirror symmetry between $N=4$ gauge theories
in three dimensions has been studied.
In this paper we will study  a similar mirror symmetry between  $N=2$ gauge 
theories in three
dimensions.

The $N=2$ supersymmetry algebra in three dimensions is only invariant under one
$U(1)_R$ symmetry. Therefore, {\it a priori} there is no notion of
mirror symmetry which, as in $N=2$ in two dimensions and $N=4$ in 
three dimensions, exchanges two commuting R-symmetries acting differently
on the supercharges. However,
there are theories having extra global symmetries commuting with the
supercharges. If we combine the $U(1)_R$ symmetry with these global symmetries,
we may be able to have two $U(1)_R$ symmetries which act
differently on the moduli space of vacua, and we can
introduce a notion of mirror 
symmetry
under which these two $U(1)_R$'s are exchanged.
One purpose of this paper is to realize this by the use of the brane 
configuration of
\cite{EGK} and application of type IIB $SL(2,\Z)$ duality following \cite{HW}.

The $N=2$ gauge theories can be constructed as the dimensional
reduction of $N=1$ gauge theories in four dimensions.
The bosonic part of the
$N=2$ vector multiplet contains the three dimensional gauge field $A_{\mu}$
and a real scalar $\varphi$ which corresponds to the $A_4$
component of the four dimensional gauge field. 
The action contains the term ${\rm Tr}[A_{\mu},A_{\nu}]^2$
and the term ${\rm Tr}[A_{\mu},\varphi]^2$.
Thus, the low energy effective theory is described by
the abelian theory in which
$\varphi$ and $A_{\mu}$ belong to a common Cartan sub-algebra
of the gauge group.
In three dimensions the photon $A_{\mu}$ is dual to a scalar field
$\sigma$. The vacuum expectation values of the scalars $\varphi$ and $\sigma$
parameterize the Coulomb branch of the theory.
The Coulomb branch has real dimension $2r$ where $r$ is
the rank of the gauge group.
Due to $N=2$ supersymmetry it is a
K\"ahler manifold of complex dimension $r$.
Note that this differs from $N=1$ theories in four dimensions where
there are no scalars in the vector multiplet and therefore there is no Coulomb 
branch.
Matter fields are in the $N=2$ chiral multiplet.
The scalar fields in the multiplet
parameterize the Higgs branch, which is determined by
the D- and F-term equations. By $N=2$ supersymmetry,
it is also a K\"ahler manifold.

In addition to the R-symmetry, the system possesses flavor symmetries depending 
on
the matter content. Also, there are global symmetries corresponding to the 
shifts of
$\sigma$ by a constant.
In non-abelian gauge theories, some of these symmetries
can be broken by instanton configuration,
but some combinations remain exact (although they might be spontaneously 
broken \cite{AHW}).
Taking suitable combinations of $U(1)_R$ and other global symmetries,
we can get two $U(1)$ R-symmetries with respect to which
we can divide the moduli space into two parts. 
The two parts will still be called the Coulomb and the Higgs branch,
and in a pair of mirror
symmetric gauge theories it is these two parts that are exchanged.
The mass parameters that exist already in four dimensions
are complex and charged with respect to a
$U(1)_R$ symmetry, while the FI parameters are real and do not carry a $U(1)_R$ 
charge.
Therefore, unlike the
$N=4$ case we should not expect a naive exchange of these parameters.
Indeed, as we will see there are many more parameters so that a
precise map between the parameters of the mirror theories is possible.

In $N=2$ supersymmetric non-abelian gauge theory in three dimensions,
a superpotential may be generated both for the Higgs and the
Coulomb branches by instantons which in three dimensions are BPS monopoles.
We will study this from the viewpoint of string theory
 by considering 
 open D-string instantons.

\begin{figure}
\begin{center}
$\mbox{\epsfig{figure=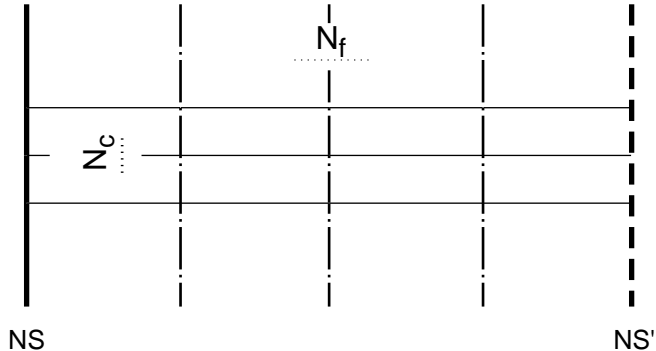}}$
\end{center}
\caption{Brane configuration for an $N=2$ $U(N_c)$ gauge theory with $N_f$ 
flavors
 in three dimensions.}
\end{figure}

We will mainly consider two brane setups
in type IIB string theory for studying the $N=2$ theories.
The first configuration is depicted in figure 1.
It consists of  one NS 5-brane whose worldvolume
has the coordinates $(x^0x^1x^2x^3x^4x^5)$, one NS${}^{\prime}$ 5-brane
whose worldvolume
has the coordinates $(x^0x^1x^2x^3x^8x^9)$,
$N_c$ D3 branes stretching between them
in the $x^6$ direction with worldvolume $(x^0x^1x^2x^6)$  and
$N_f$ D5 branes with worldvolume $(x^0x^1x^2x^7x^8x^9)$.
This configuration preserves $\frac{1}{8}$ of the $32$ space-time 
supersymmetries.
We consider the supersymmetric gauge theory obtained as the long distance limit
of the worldvolume dynamics of the D3 brane.
This is an $N=2$ supersymmetric gauge theory with $U(N_c)$ gauge
group and $N_f$ pairs of chiral multiplets in the (anti-)fundamental 
of the gauge group.

\begin{figure}
\begin{center}
$\mbox{\epsfig{figure=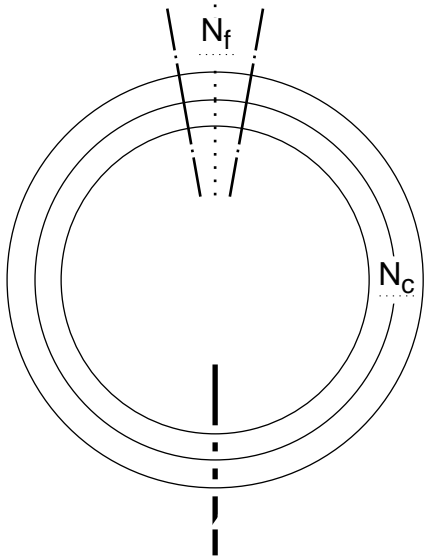}}$
\end{center}
\caption{Brane configuration for an $N=2$ 
$U(N_c)$ gauge theory with $N_f$  flavors and  three  adjoints
in three dimensions.}
\end{figure}

The second configuration is depicted in figure 2.
It consists of  an NS 5-brane  whose worldvolume
has the coordinates $(x^0x^1x^2x^3x^8x^9)$, $N_c$ D3 branes stretching 
in the $x^6$ direction with worldvolume $(x^0x^1x^2x^6)$, and
$N_f$ D5 branes with worldvolume at coordinates $(x^0x^1x^2x^7x^8x^9)$.
The $x^6$ coordinate is compactified on a circle.
This configuration also preserves $\frac{1}{8}$ of the $32$ space-time 
supersymmetries.
We consider the supersymmetric gauge theory on the worldvolume of the D3 brane
with coordinates $(x^0x^1x^2)$. This is an $N=2$ supersymmetric gauge theory 
with $U(N_c)$ gauge
group, $N_f$ pairs of chiral multiplets in the (anti-)fundamental
representation of the
gauge group and three chiral multiplets in the adjoint representation. 
The two extra pairs of massless chiral multiplets in the adjoint representation 
arise from
the compactification  of the $x^6$ coordinate on a circle. 
The field content falls into representations 
of $N=4$ supersymmetry. However half of this 
supersymmetry
is broken by the superpotential.

\section{Mirror Symmetry in Abelian $N=2$ Theories}

\subsection{A-Model}

Consider the brane configuration of figure 1 with one D3 brane ($N_c=1$ case).
In the long distance limit,
the worldvolume of the D3 brane describes an $N=2$ 
$U(1)$ gauge theory with $N_f$ 
pairs of chiral multiplets of charges $+1,-1$.
The brane configuration is invariant under rotations
in the $(x^4,x^5)$ and $(x^8,x^9)$ directions and these
correspond to the global symmetries $U(1)_{4,5}$ and 
$U(1)_{8,9}$ of the three dimensional gauge theory.

The light fields on the D3 brane worldvolume are:

\medskip
\noindent
$\bullet$ An open string ending on the D3 brane yields 
an $N=8$ $U(1)$
vector multiplet.  Only one of the seven   scalars of the multiplet 
remains after imposing the 
boundary condition at the NS and NS${}^{\prime}$ ends. Therefore 
only
an $N=2$ $U(1)$ vector multiplet remains.
The value  $x^3({\rm D3})$ (the $x^3$-coordinate of the
D3 brane) corresponds to the scalar
field $\varphi$ in the vector multiplet.
The vector  $A_{0,1,2}$ is 
dual
to the scalar field $\sigma$.
These scalars are singlets under both $U(1)$'s.
The fermions carry charge $\pm 1$ under both $U(1)_{4,5}$
and $U(1)_{8,9}$ \footnote{
We assign charge
$\pm 1$ to the spin representation of $SO(2)\cong U(1)$ and 
$\pm 2$ to the
vector representation.}.

\medskip
\noindent
$\bullet$ Open strings ending on the D3 brane and the 
$i$-th D5 brane
yield chiral multiplets $Q_{i}$ and $\tilQ_i$ of 
opposite charges
$+1$ and $-1$. The scalar components are singlet under 
$U(1)_{4,5}$
but carry charge $+1$ of $U(1)_{8,9}$. The fermions 
carry charge
$-1$ of $U(1)_{4,5}$ and  are singlets under $U(1)_{8,9}$.

\bigskip
The positions of NS, NS${}^{\prime}$ and the  D5 brane
correspond to parameters
of the three dimensional field theory.

\medskip
\noindent
$\bullet$ The difference $x^7(\NS)-x^7(\NS^{\prime})$
of the positions of the NS and $\NS^{\prime}$ 5-branes in 
the $x^7$ direction
is the Fayet-Iliopoulos parameter $\zeta^{\bf r}$
of the $U(1)$ gauge theory.
It is invariant under  the global $U(1)$ symmetries.

\medskip
\noindent
$\bullet$ The position $x^3(\D 5_i)$ of the $i$-th D5 brane 
in the $x^3$ direction is the real mass parameter $m^{\bf 
r}_i$ of the
$i$-th quark. A real mass can be considered as a scalar 
component
of $N=2$ vector multiplet of the flavor symmetry group 
$U(N_f)$.
It is invariant under the global $U(1)$ symmetries.
Note that the center of mass $\sum_{i=1}^{N_f}m_i^{\bf 
r}$
can be absorbed by a shift of $\varphi$ and is not a 
physical parameter.

\medskip
\noindent
$\bullet$ The position $x^{4,5}(\D 5_i)$ of the $i$-th D5 brane 
in the $x^4,x^5$  is the complex mass 
parameter
$m_i$ of the $i$-th quark which is present already in the
four dimensional
$N=1$ theory. Since it transforms in the vector representation 
of $SO(2)_{4,5}$
it carries charge $2$ of $U(1)_{4,5}$ and is singlet 
under $U(1)_{8,9}$.

To summarize, we list the fields and parameters of the 
gauge theory with
their
transformation properties under the global symmetry 
group
$U(1)_{4,5}\times U(1)_{8,9}$. Note that both $U(1)$'s can 
be considered as
R-symmetry groups acting on a chiral superfield $\Phi(\theta)$ 
as
$\Phi(\theta)\to\e^{Q_{\Phi}i\alpha}\Phi(\e^{-i\alpha}\theta)$. 

\beq
\begin{array}{cccl}
&U(1)_{4,5}&U(1)_{8,9}&\\
\varphi,\sigma&0&0&\\
Q_i&0&1&\,\,i=1,\ldots,N_f\\
\tilQ_i&0&1&\,\,i=1,\ldots,N_f\\
\zeta^{\bf r}&0&0&\\
m_i^{\bf r}&0&0&\,\,i=1,\ldots,N_f\\
m_i&2&0&\,\,i=1,\ldots,N_f
\end{array}
\eeq

The theory has the tree level superpotential
\beq
W=\sum_{i=1}^{N_f} m_i\tilQ_iQ_i.
\eeq
Note that $U(1)_{45}$ is broken explicitly by the complex mass parameter.

As we mentioned in the introduction  
 there is  
another
global $U(1)$ symmetry acting only on $\sigma$:
\beq
\sigma\longmapsto \sigma\,+\,\,{\rm constant}.
\eeq
The flavor symmetry group $SU(N_f)\times SU(N_f)$ is
broken to  $U(1)^{N_f-1}\times U(1)^{N_f-1}$
by the diagonal mass matrix.
Some of these symmetries are invisible in the brane 
configuration. Notice that the (independent)
flavor symmetry group
is not $U(N_f)\times U(N_f)$, because the central
$U(1)^2$ are already present in the theory as
the $U(1)$ gauge symmetry
and the axial part of $U(1)_{4,5}\times U(1)_{8,9}$.

In the abelian gauge theory in three dimensions,
there are neither perturbative nor non-perturbative  effects that
 break any of
these $2N_f+1$ $U(1)$ global symmetries.
These are exact symmetries of the quantum theory,
although spontaneous symmetry  breaking is possible \cite{AHW}.

 \subsection{B-Model}
 
Let us now perform an $SL(2,\Z)$ transformation on the above 
configuration.
Before the application of it,
we move the NS 5-brane to the right in the $x^6$
direction, crossing one D5 brane. After an $SL(2,\Z)$ transformation
we end up with the  configuration  of figure 3.

\begin{figure}
\begin{center}
$\mbox{\epsfig{figure=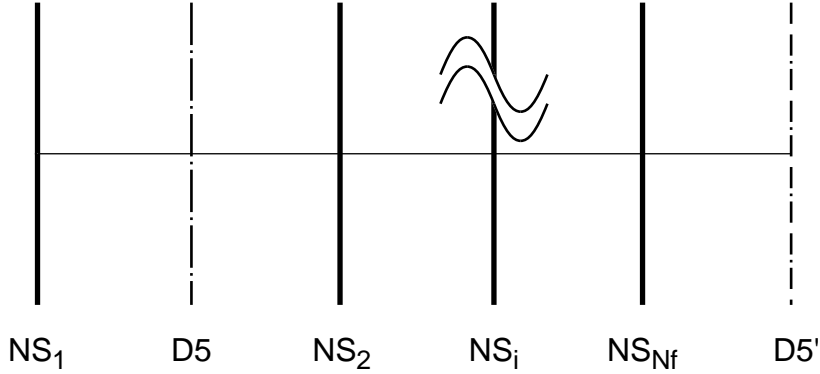}}$
\end{center}
\caption{$SL(2,Z)$ of the brane configuration
of $N=2$ $U(1)$ gauge theory with $N_f$ flavours.}
\end{figure}

The light fields on the D3 brane worldvolume are :

\medskip
\noindent
$\bullet$ An open string ending on the D3 brane
stretched between
the $\NS_i$ and $\NS_{i+1}$ 5-branes yields an
$N=4$ $U(1)$ vector multiplet,
consisting of an $N=2$ $U(1)$ vector multiplet $W_i$ and
a neutral chiral multiplet $\Phi_i$.
The value $x^7(D3)$ and vector field $A_{0,1,2}$ 
corresponds to the
scalars $\varphi_i$, $\sigma_i$ of $W_i$
while the values $x^{8,9}(D3)$ combine into the complex 
scalar
of $\Phi_i$.

\medskip
\noindent
$\bullet$ An open string ending on the D3 brane 
stretched between
the $\NS_{N_f}$ and $\D^{\prime}$ 5-branes yields an
$N=2$ neutral chiral multiplet $M$. The values 
$x^{8,9}(D3)$ correspond to
the complex scalar component of $M$.

\medskip
\noindent
$\bullet$ Open strings ending on the D5 brane and the 
D3 brane
which is stretched between the $\NS_1$ and $\NS_2$ 5-branes
yield an $N=4$ hypermultiplet $Q$, $\tilQ$ charged 
under the first
$U(1)_1$ of the $N=4$ vector multiplet $(W_1,\Phi_1)$.
The hypermultiplet is 
coupled in an $N=4$ supersymmetric way, i.e. with
a nonzero superpotential involving $Q,\tilQ,\Phi_1$
from the $N=2$ point of view. 

\medskip
\noindent
$\bullet$ Open strings ending on the D3 branes in 
the $i$-th and
$i+1$-th interval yield an $N=4$ hypermultiplet
charged as $(+1,-1)$ under the $i$-th and $i+1$-th
gauge group $U(1)_i\times U(1)_{i+1}$.
For $i=1,\ldots,N_f-2$, we denote it by
$b_{i,i+1}$, $b_{i+1,i}$ and for $i=N_f-1$, we denote it 
by
$\tilq$, $q$.

The $N=2$ chiral multiplet $M$ (``meson'') can 
be interpreted as a remnant
of an $N=4$ vector multiplet of the $U(1)$ flavor group
rotating $q$, $\tilq$, and  there is a term $M\tilq 
q$
in the superpotential.
Thus we have an $N=4$ supersymmetric $U(1)^{N_f-1}$
gauge theory broken to $N=2$
via coupling to a neutral $N=2$ chiral multiplet by 
superpotential.
This is illustrated in figure 4.

\begin{figure}
\begin{center}
$\mbox{\epsfig{figure=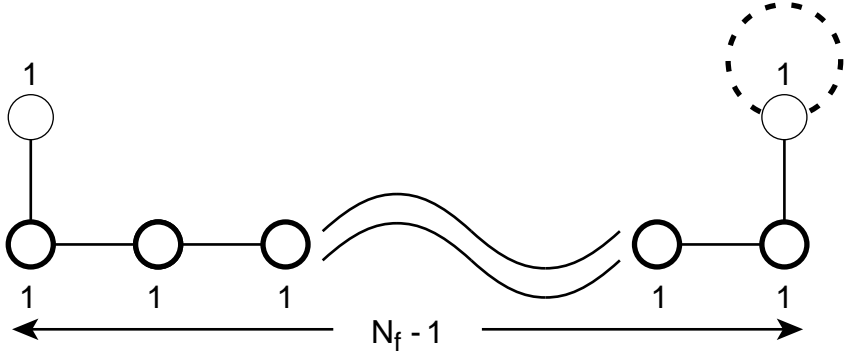}}$
\end{center}
\caption{Quiver like diagram for the B-model. The dashed circle correspond to 
the meson.}
\end{figure}

The positions of NS D and D${}^{\prime}$ 5-branes 
are  parameters of the
field theory.

\medskip
\noindent
$\bullet$ The difference 
$x^{3,4,5}(\NS_i)-x^{3,4,5}(\NS_{i+1})$
of positions of the $i$-th and $i+1$-th  NS 5-branes in 
the $x^3x^4x^5$  is the Fayet-Iliopoulos parameter 
$\zeta^{3,4,5}_i$
of the $N=4$ vector multiplet $(W_i,\Phi_i)$.
We denote $\zeta^3_i=\zeta^{\bf r}_i$ and $\zeta^{4,5}_i
=\zeta_i$.

\medskip
\noindent
$\bullet$ If the difference 
$x^{4,5}(\NS_{N_f})-x^{4,5}(\D^{\prime}5)$
is non-zero, there cannot be a D3 brane stretched in 
the interval
$\NS_{N_f}$-$\D^{\prime}5$.
Since the meson $M$ created by the D3 brane is
a remnant of the $N=4$ flavor $U(1)$ vector multiplet,
the difference is an analog of the Fayet-Iliopoulos 
parameter $\zeta_{M}$  
which enters in the superpotential as $\zeta_M M$.

\medskip
\noindent
$\bullet$ The difference $x^7(\D 5)-x^7(\D^{\prime}5)$ 
forbids
a Higgs branch corresponding to
a D3 brane stretched between the D and $\D^{\prime}$ 
5-branes, and
 it can be interpreted as the
real bare mass $m_q^{\bf r}$ of the quark $q,\tilq$.
Note that other mass parameters can be absorbed
by the shift of $\varphi_i$, $\Phi_i$ and $M$.

\medskip
To summarize, we list the fields and the parameters of the 
gauge theory. Notice that the number of parameters is
$3N_f$, the same as in the A-model.
As in the previous case, we can read the R-charges
of the two $U(1)$ symmetry groups from the brane
configuration. 

\beq
\begin{array}{cccl}
&U(1)_{4,5}&U(1)_{8,9}&\\
\varphi_i,\sigma_i&0&0&\,\,i=1,\ldots,N_f-1\\
\Phi_i&0&2&\,\,i=1,\ldots,N_f-1\\
M&0&2&\\
Q,\tilQ&1&0&\\
b_{i,i+1},b_{i+1,i}&1&0&\,\,i=1,\ldots,N_f-2\\
q,\tilq&1&0&\\
\zeta^{\bf r}_i,\zeta_i&0\oplus 2&0\oplus 
0&\,\,i=1,\ldots,N_f-1\\
\zeta_M&2&0&\\
m_q^{\bf r}&0&0&
\end{array}
\eeq
The theory possesses the tree level superpotential
\beq
W=W_{N=4}+M\tilq q-\zeta_MM,
\eeq
where $W_{N=4}$ is the superpotential of the $N=4$ 
sector
\beq
W_{N=4}=\sum_{i=1}^{N_f-1}\Phi_i
\Bigl(b_{i,i-1}b_{i-1,i}-b_{i,i+1}b_{i+1,i}-\zeta_i\Bigr
)
\label{pot}
\eeq
in which we denoted $Q,\tilQ=b_{1,0},b_{0,1}$ and
$q,\tilq=b_{N_f,N_f-1},b_{N_f-1,N_f}$.
In addition to $U(1)_{4,5}\times U(1)_{8,9}$,
we have
global symmetry group $U(1)^{N_f}$ acting as
$(q,\tilq)\mapsto (\e^{i\beta_0}q,\e^{-i\beta_0}\tilq)$;
$\sigma_i\mapsto\sigma_i+\beta_i$;
($i=1,\ldots,N_f-1$).
These $N_f+2$ global $U(1)$ symmetries are exact
in the quantum theory,
although they might be spontaneously broken.
Note that we seem to be missing global $U(1)^{N_f-1}$ 
compared with the A-model.
This is a symmetry that arises quantum mechanically
and is not manifest in the classical Lagrangian.
A similar phenomenon has been found and explained in \cite{si}
for $N=4$ supersymmetric theories.

\subsection{Verification Of The Mirror Symmetry}

\begin{figure}
\begin{center}
$\mbox{\epsfig{figure=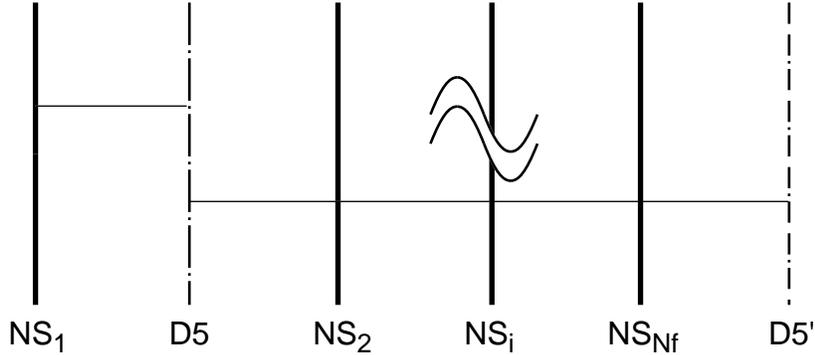}}$
\end{center}
\caption{The Higgs branch of the B-model.}
\end{figure}

There are two phases in brane 
configurations corresponding to both of the A and B models.
In the configuration of A-model, one phase corresponds to
the D3 brane  ending  on NS and NS${}^{\prime}$ 
5-branes and moving
in the $x^3$ direction (Coulomb branch), while in the second phase it is 
broken into
$N_f+1$ pieces by letting the D3 branes
end on D5 branes and the 
pieces move
in the $(x^7),x^8,x^9$ directions (Higgs branch). 
In the configuration of B-model,
in the first phase the $N_f+1$ pieces of D3 branes end on 
NS and NS or
NS and D${}^{\prime}$ 5-branes
and move in $(x^7), x^8,x^9$ directions (Coulomb branch), while in the 
second phase
the D3 brane is broken into two at the D5 brane and 
one of them moves
in the $x^3$ direction ending on D and D${}^{\prime}$ 
5-branes as in figure 5 (Higgs branch).
In the next subsection we will see that this distinction
of branches is natural from the viewpoint
of the action of the $U(1)$ R-symmetries.

In $N=2$ supersymmetric field theory in three 
dimensions,
there are no loop corrections to the F-term by the 
perturbative
non-renormalization theorem \cite{AHW}.
Moreover, in abelian gauge theory,
we expect no non-perturbative effect, and
thus there is no dynamical generation
of a superpotential in these models. However, there are
loop corrections to the D-term, and thus we can at most
compare the complex structure of
the moduli spaces of vacua. 
The complex structure of Higgs branch is determined
at the classical level, and is expected to be unchanged
by quantum corrections.
However, the Coulomb branch is obtained after a duality
transformation, and the global structure may be changed in
the dual variables. Nevertheless, we expect that,
like in the $N=4$ case \cite{SW}, the global and complex
structure can be
determined by looking at the behavior at infinity of
the moduli space, and that a one-loop analysis is
sufficient to determine them.      
It turns out that the one loop Coulomb branch does have a natural
complex structure, and as in the case of Higgs branch
it will not be further corrected. Thus, we compare the complex
 structure of one loop corrected Coulomb branch of
the A-model and the classical Higgs branch of the B-model.

\subsubsection{ ${\cal M}_C(\mbox{A-Model}) = {\cal M}_H(\mbox{B-Model})$}

A Coulomb branch of the A-model is possible only when 
the FI parameter
$\zeta^{\bf r}$ vanishes, while a Higgs branch of the 
B-model
is possible only when the real bare mass $m_q^{\bf r}$ is 
zero.
Therefore, we can identify $\zeta^{\bf r}=m_q^{\bf r}$.

We first consider the one loop metric of the Coulomb 
branch of the
A-model. Now it is useful to consider the model --- 
$N=2$ supersymmetric
QED with $N_f$ electrons --- as ``embedded'' in the 
corresponding $N=4$
model whose mass 3-vector $\vec{m}_i$ is given by
$(m_i^{\bf r},m_i)$.
The only difference in the one loop computation of 
the Coulomb branch metric
is that the $N=4$ vector multiplet contains
a neutral chiral multiplet $\Phi$
which provides the center of complex bare mass of the 
electrons.
Thus, the one loop metric of the $N=2$ Coulomb branch
is given by the one loop metric of the $N=4$ Coulomb 
branch
{\it restricted to} the hyperplane
$\Phi=m_{\rm c}:=\sum_im_i/N_f$. In \cite{dhoo}, the one 
loop metric on the 
$N=4$ Coulomb branch has been  computed.
In the limit where the bare coupling $1/e^2\to 0$, it is the ALE space of 
$A_{N_f-1}$ type
whose $\zeta$ parameters are given by 
$\vec{m}_i-\vec{m}_{i-1}$
($i=1,\ldots,N_f$; $\vec{m}_0:=\vec{m}_{N_f-1}$).
Thus, the Coulomb branch of the A-model is given by
\beq
\Bigl\{\,\,\Phi=m_{\rm c}\,\,\Bigr\}\subset
\mbox{ALE}_{A_{N_f-1}}\bigl(\,\vec{m}_i-\vec{m}_{i-1}\,
\bigr).
\eeq
The limit $1/e^2\to 0$ was also taken while studying the 
mirror symmetry in $N=4$ theories and corresponds to taking the IR
limit.

It is also useful to note that the $N=4$ sector of the 
B-model is the one first
considered in \cite{si} as the mirror of $N=4$ QED 
with $N_f$ electrons.
Since the model is obtained by coupling it to the 
$N=2$ meson sector
via the superpotential (\ref{pot}),
the only difference in the Higgs branch is that there is an 
extra constraint
$\partial W/\partial M=0$. The $N=4$ Higgs branch is 
again the ALE space
whose $\zeta$ parameters are given by
$\vec{\zeta}_i=\zeta_i^{3,4,5}$ 
($i=0,1,\ldots,N_f-1$;
$\vec{\zeta}_{0}:=-\sum_i\vec{\zeta}_i$). Thus, the 
$N=2$ Higgs branch is given by
\beq
\Bigl\{\,\,\tilq q=\zeta_M\,\,\Bigr\}
\subset 
\mbox{ALE}_{A_{N_f-1}}\bigl(\,\,\vec{\zeta}_i\,\,\bigr).
\eeq

That these two have the same complex structure under a 
certain relation of
$\vec{m}_i$ and $\vec{\zeta}_i$, $\zeta_M$ can be seen 
by looking at the
one loop computation given in Section 4 of \cite{dhoo}.
In fact, they are the same submanifold of the same 
hyper-K\"ahler
manifold whose defining equation is holomorphic
with respect to one distinguished complex structure.
To illustrate this,
let us consider the case in which all the bare masses are 
the same
($m_i=m_{\rm c}$) in the A-model
and all the FI parameters are vanishing 
($\vec{\zeta}_i=0$)
in the B-model.
In this case, both ALE spaces coincide with the orbifold 
$\C^2/\Z_{N_f}$.
Let $(z_1,z_2)$ be the coordinates of $\C^2$ on which 
$\Z_{N_f}$ acts by
$(z_1,z_2)\mapsto (\e^{2\pi i l/N_f}z_1,\e^{-2\pi i 
l/N_f}z_2)$.
In the A-model, $\varphi$, $\sigma$, $\Phi$ are 
expressed in terms of
$z_1, z_2$ by $\varphi=|z_1|^2-|z_2|^2$,
$\sigma=2{\rm arg}(z_1)$, and
$\Phi=z_1z_2$. By introducing the $\Z_{N_f}$ invariant 
coordinates,
$x=z_1^{N_f}$, $y=z_2^{N_f}$
the Coulomb branch of the A-model is described by
\beq
xy=m_{\rm c}^{N_f}\,.
\label{CoulombA}
\eeq

The Higgs branch of the B-model
is determined in this case by the equations
$|b_{i,i+1}|=|\tilQ|=|\tilq|$,
$|b_{i+1,i}|=|Q|=|q|$,
$b_{i,i+1}b_{i+1,i}=\tilQ Q=\tilq q$,
and $\tilq q=\zeta_M$
modulo gauge transformations.
Using gauge invariant coordinates
$x=Qq\prod_{i=1}^{N_f-2}b_{i+1,i}$, $y=\tilQ\tilq 
\prod_{i=1}^{N_f-2} b_{i,i+1}$,
it is expressed as
\beq
xy=\zeta_M^{N_f},
\eeq
which is the same as 
(\ref{CoulombA}) if we identify $m_{\rm c}=\zeta_M$.

\subsubsection{ ${\cal M}_H(\mbox{A-Model}) =
{\cal M}_C(\mbox{B-Model})$}

When we turn off all the mass parameters of the  
A-model and set
the vev of the scalars in the vector multiplet to zero, we get a 
Higgs branch of
maximal dimension in which $Q_i$ and $\tilQ_i$ are 
turned on.
The classical D-term equation is given by
\beq
\sum_{i=1}^{N_f}|Q_i|^2-\sum_{i=1}^{N_f}|\tilQ_i|^2=\zeta^{\bf r}.
\eeq
The Higgs branch is obtained as the $U(1)$ quotient of 
the set of
solutions. This is the standard K\"ahler quotient of 
$(\C\oplus
\C^{\vee})^{\oplus N_f}$ by the $U(1)$ action. As a 
complex manifold,
it is obtained as the quotient of 
$(\C\oplus \C^{\vee})^{\oplus N_f}$
by the $\C^{\times}$ action, in which we throw away some
``bad orbits'' depending on the value of $\zeta^{\bf 
r}$. (A good explanation can be found in \cite{Witten}.)
For $\zeta^{\bf r}\ne 0$, it is isomorphic to the total 
space
of the direct sum of $N_f$ tautological bundles ${\cal 
O}(-1)^{\oplus N_f}$
over $\C{\bf P}^{N_f-1}$. For $\zeta^{\bf r}=0$, it is 
a singular space which is described by 
the quadratic relations
$x_{i,k}x_{j,l}=x_{i,l}x_{j,k}$ 
where $x_{i,j}$ are the gauge invariant
coordinates $x_{i,j}=Q_i\tilQ_j$.

\medskip
For the B-model, when we turn off all the FI 
parameters and set the vev of 
$Q, \tilQ,b_{i,i\pm 1}, q,\tilq$ to zero, we obtain a Coulomb 
branch
parameterized by  $\varphi_i,\sigma_i,\Phi_i,M$.
Recall that the $N=4$
sector of the model is the model  considered in 
\cite{si}
as a mirror of $N=4$ QED with
$N_f$ electrons. Note that the meson $M$ provides the 
complex part
of the mass $\vec{m}=(m_q^{\bf r},M)$ of the field 
$q,\tilq$.
As computed in \cite{dhoo}, the one loop corrected Coulomb branch 
of
the $N=4$ model is the same as the classical Higgs 
branch of its mirror,
$N=4$ QED with $N_f$ electrons, with its FI parameter 
given by $\vec{m}$.
It is the set of $U(1)$ orbits of $(Q_i,\tilQ_i)$
satisfying
\beqa
&&\sum_{i=1}^{N_f}|Q_i|^2-\sum_{i=1}^{N_f}|\tilQ_i|^2=m_
q^{\bf r},\\
&&\sum_{i=1}^{N_f}\tilQ_iQ_i=M.
\eeqa
In the  $N=2$ Coulomb branch  $M$ is free 
to vary.
This means that the second equation does not restrict the vev of $\tilQ_i,Q_i$, 
and the Coulomb 
branch of the B-model
is the same as the Higgs branch of the A-model, provided 
$\zeta^{\bf r}=m_q^{\bf r}$.

\subsubsection{ The Action of $U(1)$ R-Symmetry Groups}

Recall that we have two $U(1)$ R-symmetries
$U(1)_{4,5}$ and $U(1)_{8,9}$ coming from the invariance 
of the brane configuration.
In this subsection, we shall see that these two $U(1)$'s,
modified by combination with other global symmetries,
act on Coulomb or Higgs branches, one group on one 
branch,
the other on the other,
and that these actions are interchanged under mirror 
symmetry.

In the A-model, we define $U(1)_H=U(1)_{8,9}$ and
$U(1)_C$ as the diagonal subgroup
of the product of
$U(1)_{4,5}$ and the global $U(1)$ acting as shifts of
$\sigma$.
$U(1)_C$ acts on the coordinates and
parameters of the Coulomb branch as
$\sigma\mapsto \sigma+2\alpha$, $m_i\mapsto 
\e^{2i\alpha}m_i$,
and $U(1)_H$ acts on the Higgs branch as
$Q_i,\tilQ_i\mapsto 
\e^{i\alpha}Q_i,\e^{i\alpha}\tilQ_i$.

In the B-model, we define $U(1)_H=U(1)_{4,5}$ and
$U(1)_C$
as the diagonal of the product of
$U(1)_{8,9}$ and the global $U(1)$ acting as shifts of 
$\sigma_i$.
$U(1)_C$ acts on the Coulomb branch as
$\sigma_i\mapsto \sigma_i+2\alpha$,
$(\Phi_i,M)\mapsto (\e^{2i\alpha}\Phi_i, \e^{2i\alpha}M)$,
while $U(1)_H$ acts on the Higgs branch
as multiplication by $\e^{i\alpha}$ on
$Q,\tilQ,b_{i,i\pm 1},q,\tilq$ and multiplication
by $\e^{2i\alpha}$ on $\zeta_i, \zeta_M$.

To see that $U(1)_C$ and $U(1)_H$ of the A-model
is mapped to $U(1)_H$ and $U(1)_C$ of the B-model,
we recall the transformation mapping two $SU(2)$ doublet 
${\bf 2} \times {\bf 2}$
to ${\bf 1}\oplus {\bf 3}$, which was used in the
comparison of the Coulomb and Higgs 
branches of the 
$N=4$ mirror theories\cite{dhoo}.

Let $(z_1,\overline{z_2})$ be the coordinates of the 
$SU(2)$
doublets. In terms of the quaternion coordinate
${\bf q}=z_1+z_2j$,
the $SU(2)$ action is the left multiplication by unit 
quaternions
$Sp(1)\subset {\bf H}^{\times}$.
We define $\sigma$ and 
$\vec{\varphi}=(\varphi^1,\varphi^2,\varphi^3)$ by
${\bf q}=a\e^{i\sigma/2}$; $\overline{a}=-a$, and
${\bf q}i\overline{\bf q}=\varphi^1i+\varphi^2j+\varphi^3k$.
Under an $SU(2)$ rotation, $\vec{\varphi}$ transforms as 
a vector ${\bf 3}$.
With respect to the $U(1)$ subgroup 
$\{\e^{i\alpha}\}\subset Sp(1)$,
$z_1,z_2$ transform as 
$\e^{i\alpha}z_1,\e^{i\alpha}z_2$,
while $\sigma$, $\varphi^1=|z_1|^2-|z_2|^2$, and
$\Phi:=\varphi^2+i\varphi^3=-2iz_1z_2$ transform as 
$\sigma\mapsto
\sigma+2\alpha$, $\varphi^1\mapsto\varphi^1$ and $\Phi\mapsto 
\e^{2i\alpha}\Phi$.

We now observe that these transformations are 
indeed mapped into each other
under the identification 
of the Coulomb branch of the A-model with
the Higgs branch of the B-model, and vice versa.

\section{Mirror Symmetry in Non-Abelian $N=2$ Theories}

\subsection{A-Model}
Consider the brane configuration of figure 1 with $N_c$ D3 branes.
In the long distance limit,
the worldvolume of the D3 brane describes
an $N=2$ 
 $U(N_c)$ gauge theory with $N_f$ 
pairs of chiral multiplets.
As discussed in the previous section
the brane configuration is invariant under rotation
in the $(x^4,x^5)$ and $(x^8,x^9)$ directions and these
correspond to global symmetries $U(1)_{4,5}$ and 
$U(1)_{8,9}$ of the three dimensional gauge theory.

As in the previous section we read from the brane configuration the fields and 
parameters of the
gauge theory on the worldvolume of the D3 brane, and their  
transformation properties under the global symmetry 
group
$U(1)_{4,5}\times U(1)_{8,9}$. 
This is summarized in the following.

\beq
\begin{array}{cccl}
&U(1)_{4,5}&U(1)_{8,9}&\\
\varphi_a,\sigma_a&0&0&\,\,a=1,\ldots,N_c\\
Q_i&0&1&\,\,i=1,\ldots,N_f\\
\tilQ_i&0&1&\,\,i=1,\ldots,N_f\\
\zeta^{\bf r}&0&0&\\
m_i^{\bf r}&0&0&\,\,i=1,\ldots,N_f\\
m_i&2&0&\,\,i=1,\ldots,N_f
\end{array}
\label{list}
\eeq

\subsection{B-Model}

Let us now perform an $SL(2,\Z)$ transformation on the above 
configuration.
Before the application of it,
we move the NS 5-brane to the right in 
the $x^6$
direction, crossing $N_c$  D5 branes.
We end up with the  configuration  of figure 6.

\begin{figure}
\begin{center}
$\mbox{\epsfig{figure=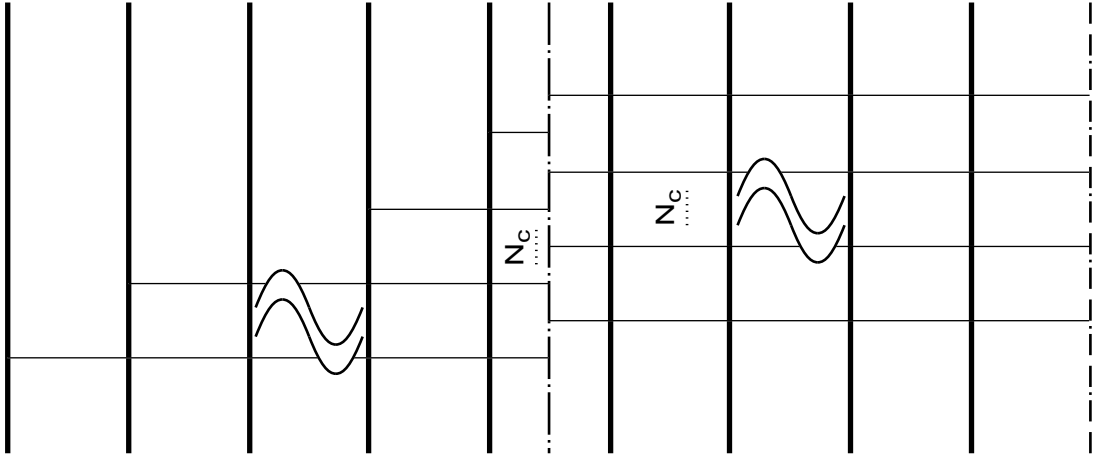}}$
\end{center}
\caption{Brane configuration of the B-model.}
\end{figure}

The gauge group of the B-model is $U(N_c)^{N_f-N_c}\prod_{i=1}^{N_c-1}U(i)$.
For $N_c=1$ it reduces to the abelian duality of the previous section with 
gauge
group $U(1)^{N_f-1}$.

Again we summarize the fields  and parameters of the
gauge theory on the worldvolume of the D3 brane, and their  
transformation properties under the global symmetry groups.

\beq
\begin{array}{cccl}
&U(1)_{4,5}&U(1)_{8,9}&\\
\varphi_i,\sigma_i&0&0&\,\,i=1,\ldots,N_f-1\\
\Phi_i&0&2&\,\,i=1,\ldots,N_f-1\\
M_{\alpha}&0&2&\,\,\alpha=1,\ldots, N_c\\
Q,\tilQ&1&0&\\
b_{i,i+1},b_{i+1,i}&1&0&\,\,i=1,\ldots,N_f-1\\
q_{\alpha},\tilq_{\alpha}&1&0&\,\,\alpha=1,\ldots,N_c\\
\zeta^{\bf r}_i,\zeta_i&0\oplus 2&0\oplus 
0&\,\,i=1,\ldots,N_f-1\\
\zeta_M&2&0&\\
m_q^{\bf r}&0&0&
\end{array}
\eeq
We follow the notations of the abelian case. The only difference is that now 
the 
fields carry also gauge indices of the corresponding gauge group
which we abbreviated.

The $N=2$  meson  $M$  couples
to $q$, $\tilq$ by the term
$\sum M_{\alpha}\tilq_{\alpha} q_{\alpha}$
in the superpotential.
The B-model 
is an $N=4$ supersymmetric $U(N_c)^{N_f-N_c}\prod_{i=1}^{N_c-1}U(i)$
gauge theory broken to $N=2$ by coupling to the $N=2$ sector
via the superpotential, as in figure 7.

\begin{figure}
\begin{center}
$\mbox{\epsfig{figure=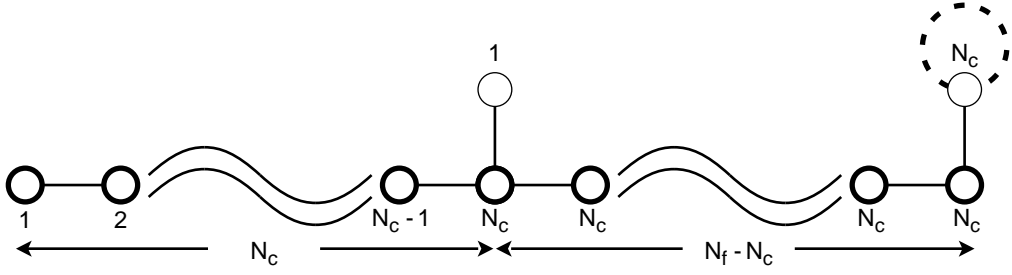}}$
\end{center}
\caption{Quiver like diagram for the B-model.}
\end{figure}

As in the Abelian case we expect that the mirror map exchanges an
appropriately defined Higgs and Coulomb branches.
However, since in the non-Abelian theories we expect in general
that superpotentials
will be generated, the analysis of the previous section has to be modified.
The full details of these modifications remain to be uncovered.
 
The $SL(2,Z)$ transformation suggests that 
 the mirror map exchanges the  FI   and mass parameters
of the A-model with the mass and FI parameters of the B-model
\beqa
&&\zeta^{\bf r} = m_q^{\bf r},
~~~~~m_i^{\bf r}-m_{i-1}^{\bf r}=\zeta_i^{\bf r},~~~~~i=1,...,N_f-1,\\
&&m_c= \zeta_M,
~~~~~ m_{i} - m_{i-1}=  \zeta_{i},~~~~~i=1,...,N_f-1,
\eeqa
where $m_c$ is the average of the complex mass parameters of the A-model.

\section{Breaking $N=4$ by a superpotential}

Mirror symmetry in three-dimensional
$N=4$ gauge theories
have been studied in \cite{dhoo,dhooy}
based on quiver diagrams.
In this section we study mirror $N=2$ gauge theories
which are obtained from these
$N=4$ by breaking half of the supersymmetry due to a vanishing superpotential.

\subsection{A-Model}

Consider the brane configuration in figure 2, with
 one NS 5-brane with coordinates $(x^0x^1x^2x^3x^8x^9)$, $N_f$ Dirichlet 5-branes 
 with coordinates $(x^0x^1x^2x^3x^4x^5)$ and 
 $N_c$ Dirichlet 3-branes with coordinates $(x^0x^1x^2x^3)$.
 The coordinate $x^3$ is compactified on a circle.
In the long distance limit,
worldvolume theory of the D3 branes is that of $N=2$ supersymmetric theory 
with $U(N_c)$ gauge group,
$N_f$ pairs of
 chiral multiplets in the fundamental and the dual
representations, and three massless chiral multiplets in the adjoint
representation.
Two extra adjoint chiral multiplets arise from the compactification of the
coordinate $x^3$ on a circle. They will be denoted by $A,\tilA$.
The $N=4$ supersymmetry is broken to $N=2$ since the superpotential
vanishes. In order to see that the Yukawa coupling in the $N=4$ superpotential 
vanishes
we list, as before, the fields, parameters and charges.

\beq
\begin{array}{cccl}
&U(1)_{4,5}&U(1)_{8,9}&\\
\varphi,\sigma &0&0&\\
\Phi &0&2&\\
Q_i,\tilQ_i&0&1&\,\,i=1,\ldots,N_f\\
A,\tilA&1&0&\\
m_i^{\bf r}&0&0&\,\,i=1,\ldots,N_f\\
m_i&2&0&\,\,i=1,\ldots,N_f
\end{array}
\eeq

Indeed the Yukawa coupling term $\tilQ \Phi Q$ does not carry the correct 
$U(1)$ charges
and therefore is absent in the superpotential, thus breaking the $N=4$ 
supersymmetry
to $N=2$.
Since we have only one NS 5-brane, the FI parameters associated with the 
$U(1)$ factor in the gauge group vanishes.
Note also that there are really only $N_f-1$  mass parameters, since one linear 
combination
can be shifted by shifting $\varphi,\Phi$.
The mass of the adjoint chiral multiplets is zero by construction.

\subsection{B-Model}

Performing an $SL(2,Z)$ on the configuration of figure 2 we get the 
configuration of figure 8.

\begin{figure}
\begin{center}
$\mbox{\epsfig{figure=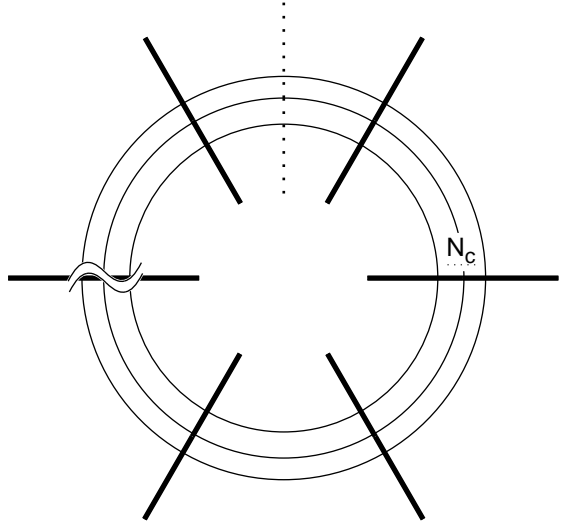}}$
\end{center}
\caption{Brane configuration of the B-model obtained by $SL(2,Z)$ 
transformation
of figure 2.}
\end{figure}

The list of the fields, parameters and charges reads now

\beq
\begin{array}{cccl}
&U(1)_{4,5}&U(1)_{8,9}&\\
\varphi_i,\sigma_i&0&0&\,\,i=1,\ldots,N_f\\
\Phi_i&0&2&\,\,i=1,\ldots,N_f\\
Q,\tilQ&0&1&\\
b_{i,i+1},b_{i+1,i}&1&0&\,\,i=1,\ldots,N_f\\
\zeta^{\bf r}_i,\zeta_i&0\oplus 2&0\oplus 
0&\,\,i=1,\ldots,N_f-1\\
\end{array}
\eeq
where by $b_{N_f,N_f+1}$ we mean $b_{N_f,1}$.
There are no mass parameters in the B-model since they can all be set to zero 
by shifting
$\varphi_i, \Phi_i$. 
Note also that there are only $N_f-1$ FI parameters since the sum of them 
vanishes
$\sum_{i=1}^{N_f}\zeta^{\bf r}_i = \sum_{i=1}^{N_f}\zeta_i = 0$.
This corresponds to the fact that the mass of the adjoint chiral multiplets in 
the A-model
is zero. The B-model has a tree level superpotential containing
terms $b\Phi b$ and terms $Q \tilde{Q} b b$. In addition, there
will in general be nonperturbative corrections to the superpotential.
The precise form of these corrections is not known, and this obstructs a
more detailed check of the mirror symmetry.

According to the $SL(2,Z)$ duality,
the mirror map between the mass parameters of the A-model and the FI parameters 
of the B-model takes the form

\beq
m_{i} - m_{i-1}=  \zeta_{i},~~~~~m_i^{\bf r}- m_{i-1}^{\bf r}=
\zeta^{\bf r}_i~~~~~~~~~~i=1,...,N_f-1.
\label{mzeta}
\eeq

The above construction of mirror pairs generalizes in a straightforward
way to all the mirror quiver diagrams of $N=4$ gauge theories that 
 were discussed in \cite{dhooy}.

\section{Comments on $N=2$ Theories}

\subsection{Duality in Three Dimensions}

In \cite{EGK} a duality between $U(N_c)$ and $U(N_f-N_c)$ $N=1$ gauge theories 
in four dimensions
has been studied. 
Upon T-duality in $x^3$, the same brane configurations can be used to study a 
similar duality
in three dimensions.
As an example,
consider the brane configuration in figure 1 with one D3 brane and two D5 
branes.
 It describes  in the coordinates
$(x^0x^1x^2)$ $N=2$  $U(1)$ gauge theory with
two pairs of chiral multiplets in the fundamental representation.
Moving the NS 5-brane through the D5 branes using \cite{HW}
and around the NS${}^{\prime}$ 5-brane
we get the configuration of figure 9, which
describes a $U(1)$ gauge theory with two pairs of chiral multiplets
 in the fundamental and the dual representations
 $(q^i,
\tilde{q}_{\tilde{i}})$ , a meson
$M_i^{\tilde{i}}$ and a superpotential $W= M_i^{\tilde{i}}q^i
\tilde{q}_{\tilde{i}}$.

\begin{figure}
\begin{center}
$\mbox{\epsfig{figure=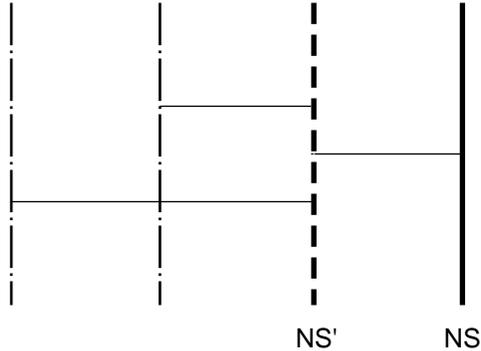}}$
\end{center}
\caption{Brane configuration for an $N=2$ $U(1)$ gauge theory with two 
flavours, a meson and
a superpotential.}
\end{figure}

The complex dimension of the Higgs branch of the theory we ended with is four
while that of the original theory is three.
The $U(1)$ gauge theories in three dimensions do not have monopoles and 
therefore there are
no instanton corrections.
Also we do not expect any strong dynamics 
that 
will generate a superpotential \cite{supot}.
It is easy to see that  a non perturbative
generation of a superpotential is also forbidden by the $U(1)$ symmetries 
(\ref{list}). 
Therefore, the classical counting of the dimensions of the Higgs branches
is valid quantum mechanically.
This suggests that the duality between $N=1$ $U(N_c)$ and $U(N_f-N_c)$ gauge 
theories
in four dimensions  \cite{EGK} is not valid in three dimensions.

\subsection{Superpotentials and Open D-String Instantons }

	It has been argued that nonperturbative dynamics of 
supersymmetric gauge theories in three dimensions is controlled by instantons 
\cite {supot}.  
In this section we will study instantons from 
string theory viewpoint and analyze their contribution to 
the superpotential.  
In three dimensions the instanton carries a magnetic charge.
The magnetic charge is 
 mediated by  the scalar dual to the photon $\sigma$ \cite {poly}. 
The instantons from the 
 string theory viewpoint are
the D-strings that end on the D3 branes\cite {HW,dhooy}.  
The boundary of 
a D-string is the worldline of a magnetic monopole in the D3 brane \cite 
{douglas}.  
To break half of the supersymmetry, it must be holomorphically embedded 
\cite {cy}, which in this case means being flat and 
orthogonally intersecting other branes.  
To qualify as an instanton configuration for the effective three dimensional 
theory, the D-string  worldvolume must be Euclidean and compact.  
Therefore it must be bounded on all sides.

\begin{figure}
\begin{center}
$\mbox{\epsfig{figure=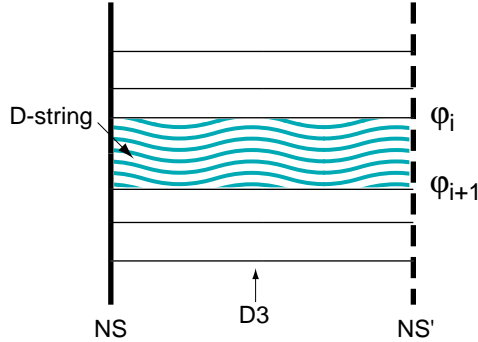}}$
\end{center}
\caption{Open D-string instanton generation of a superpotential.}
\end{figure}

	One such instanton is illustrated in figure 10, a D-string 
stretched (shaded region) between parallel pairs of D3 and NS 5-branes.  They are 
the $SL(2,Z)$ 
dual of the open string instantons of \cite {syz,cy}.
Here we consider a generic point in the Coulomb branch of 
the moduli space, so the $U(N_c)$ gauge group on the D3 worldvolume
is broken to its 
maximal Abelian subgroup.  The $\varphi$'s are the VEV's of the expectation 
values of the scalars in the vector multiplets.  They also parameterize 
the positions of the D3 branes in the $x^3$ direction.  We use the convention 
that $\varphi_i > \varphi_{i+1}$.  A monopole of charge 
$(n_1, n_2 - n_1, n_3 - n_2 \dots, n_{N_c-1} - n_{N_c-2}, - n_{N_c-1})$
with respect to the 
unbroken $U(1)$'s can be thought of as a multi-monopole solution 
made up of $n_i$ {\em fundamental monopoles} 
of the i-th type.  The latter have charges $1$ and $-1$ with respect to the i-th 
and 
(i+1)-th $U(1)$ respectively and 0 for the rest \cite{wein1}.  This has a 
straightforward 
interpretation in string theory.  The i-th fundamental open D-string 
instanton is represented by the shaded area in figure 10.  More general instanton 
configuration are just combination of this type.  

The instanton contributions to the path integral 
take the  form of $K' e^{-S_0 - i\sigma}$ \cite {poly}, where $K'$ is a factor 
that includes the one loop determinant and $S_0$ is the classical 
action for the instanton background
 $S_0 \sim - \frac \varphi {e^2}$.  
$\sigma$  is the dual to the photon of the unbroken $U(1)$.  
It emerges from field theory after summing the instantons in the
dilute instanton gas approximation \cite{poly,AHW}.
It is also expected  
by holomorphy arguments.  
All these  have a counterpart in string theory language.

Naturally, instanton corrections in string theory come in the form of 
\beq
	K e^{-S_{\rm{D-string}}}
\eeq
where K is a factor that includes the one-loop determinant of the massive fields 
on the D-string worldsheet, and $S_{\rm {D-string}}$ is the D-string worldsheet 
action.
This action 
contains two pieces:
\beq
	S_0 = S_{\rm Nambu-Goto} + i  \int_{\rm boundary}
	\tilde{A} \cdot dX,
\eeq
The Nambu-Goto action simply yields the area of the Euclidean D-string divided 
by the tension of the D-string \cite {dhooy}.  Thus
\beq
	S_{Nambu-Goto} =  {\mbox{area}} \times {\rm {tension}} = 
		\frac { [(\varphi_{i}-\varphi_{i+1}) \alpha'] 
			\times [ g_{\rm st} / {e^2}]}
			{g_{st} \alpha'} 
	= \frac {\varphi_{i} - \varphi_{i+1}} {e^2}
	\comma
\eeq
where we used the relation between the three dimensional gauge
coupling $e$, the string coupling $g_{\rm st}$ and the distance $r$ 
between the NS 5-branes in the $x^6$ direction:
$\frac{1}{e^2} = \frac{r}{g_{\rm st}}$.

In addition, there is the contribution from the boundary of the D-string.  It 
couples to the electric and magnetic gauge potential on the D5 and D3 branes 
with coupling constants $g$ of the respective theories.  
The former is not dynamical but the latter is 
important.  Denote the magnetic and electric
 gauge potentials and field strengths 
by tilded and untilded symbols respectively, then 
\beq	\label {dualph}
	\epsilon_{ijk} F_{ij} = g_{\rm st} 
		\tilde F_{k 6} = g_{\rm st} (\pa_k \tilde A_6 - \pa_6 \tilde 
A_k),
\eeq
where $i,j,k$ take value among $0, 1, 2$.  
Applying $SL(2,Z)$ to the discussion in \cite {HW},
one deduces that when a D3 brane ends on two NS 5-branes, the magnetic gauge 
field 
vanishes in the effective three dimensional theory but $\tilde A_6$ survives. 
Equation (\ref {dualph})
now reads 
\beq
	 \epsilon_{ijk} F_{ij} = g_{\rm st} \pa_k \tilde A_6.
\eeq
Thus $g_{\rm st} \tilde  A_6 = e^2 \sigma$ is the dual of the photon. 
The  
contribution of the second term in (\ref {dualph}) is now
\beq
	\int_{boundary} \tilde{A} \cdot dX 
		=   {\sigma_{i}-\sigma_{i+1}}  
\eeq
Therefore the correction from such an instanton is proportional to 
\beq
	 e^{-((\varphi_i - \varphi_{i+1})/e^2 + i (\sigma_i - \sigma_{i+1}))}
 = e^{(Z_{i+1} - Z_i)}, \;\;\;\;\;\; Z_i \equiv \frac{\varphi_i}{e^2} + i  
\sigma_i,
\eeq
in agreement with field theoretic expectation.  Note that $S_{Nambu-Goto}$ is 
insensitive to the orientation of the D-string but the $i \sigma$ term is.  
For anti-(D-string)instanton it changes sign so an anti-instanton  correction is 
anti-holomorphic.  Note also that the factor $K$ cannot 
have any dependence on the fields $Z$.

The instantons, being  BPS objects, 
 break one half of the N=2 supersymmetry.  This is consistent with 
the stringy interpretation as the D-string configuration in 
figure 10   breaks by a further half the supersymmetry preserved by the 
the NS5-D3-NS${}^{\prime}$5 configuration.  This yields  two 
zero modes.  By the Callias index theorem \cite {callias,wein1} 
the index of the Dirac operator for the 
gauginos is precisely two only for the fundamental monopoles 
considered above. 
Since we have only two fermionic zero modes the instantons correct the 
superpotential. 
Thus, only the fundamental monopoles contribute to the 
superpotential:
\beq
	W_{dyn} = K \sum_{i = 1}^{N_c-1}
		e^{Z_{i+1} - Z_i}
		\stop
\eeq
This result has been derived in \cite {KV} by considering M-theory
on a Calabi-Yau 4-fold.  

The above procedure can be applied to gauge theories that include matter.
In such cases we get extra fermionic zero modes, which in the string theory 
framework arise
from the D5 branes intersecting the D-string worldsheet at a point.

\section*{Acknowledgments}
We would like to thank D.~Kutasov and H.~Ooguri
  for discussions. This work is supported in part by 
NSF grant PHY-951497 and DOE grant DE-AC03-76SF00098. JdB is a fellow of
the Miller Institute for Basic Research in Science.

\end{document}